\documentclass[preprint,12pt]{elsarticle}

\usepackage{amssymb}

\usepackage{lineno}
\usepackage{afterpage}
\usepackage{subfigure}
\usepackage{amsmath}

\usepackage{hyperref}

\journal{Physics Letters B}

\newcommand{\QGSJET}{QGSJET-I\hspace{-.1em}I-04}
\newcommand{\EPOSLHC}{EPOS-LHC}
\newcommand{\SIBYLL}{Sibyll~2.3d}
\newcommand{\DPMJET}{DPMjet-I\hspace{-.1em}I\hspace{-.1em}I~2019.1}

\newcommand{\pp}{{\it pp} }
\newcommand{\etaregionZero}{$\eta \, > 8.5$}
\newcommand{\etaregion}[2]{$#1 \, < \eta \, < #2$}
\newcommand{\pT}{$p_{\mathrm{T}}$}
\newcommand{\xF}{$x_{\mathrm{F}}$}
\newcommand{\pizero}{$\pi^{0}$}

\begin{document}

\begin{frontmatter}



\title{Measurement of forward photon production cross-section in {\it pp} collisions at $\sqrt{s}$ = 510 GeV with RHICf detector}


\author[a,b]{O.~Adriani}
\author[a,b]{E.~Berti}
\author[b]{L.~Bonechi}
\author[a,b]{R.~D’Alessandro}
\author[c,d]{Y.~Goto}
\author[e]{B.~Hong}
\author[f]{Y.~Itow}
\author[g]{K.~Kasahara}
\author[d,h]{M.~H.~Kim}
\author[i]{Y.~Kim}
\author[j]{J.~H.~Lee}
\author[c,i]{S.~Lee}
\author[j]{T.~Ljubicic}
\author[k,l]{H.~Menjo\corref{cor1}}
\ead{menjo@isee.nagoya-u.ac.jp}
\author[c,d]{I.~Nakagawa}
\author[j]{A.~Ogawa}
\author[i]{S. Oh}
\author[k]{K.~Ohashi$^{**}$ \footnote{$^{**}$ Present address: University of Bern, Bern, CH-3012, Switzerland}}
\author[j]{R.~Pak}
\author[f]{T.~Sako}
\author[m]{N.~Sakurai}
\author[k]{K.~Sato}
\author[c,d]{R.~Seidl}
\author[n]{K.~Tanida}
\author[o]{S.~Torii}
\author[p,q,r]{A.~Tricomi}

\cortext[cor1]{Corresponding author.}

\affiliation[a]{organization={Department of Physics and Astronomy, University of Florence}, 
    city={Sesto Florentino},
    postcode={I-50019}, 
    country={Italy}}
\affiliation[b]{organization={INFN Section of Florence}, 
    city={Sesto Fiorentino},
    postcode={I-50019}, 
    country={Italy}}
\affiliation[c]{organization={RIKEN Nishina Center for Accelerator-Based Science}, 
    city={Wako},
    postcode={351-0198}, 
    state={Saitama},
    country={Japan}}
\affiliation[d]{organization={RIKEN BNL Research Center, Brookhaven National Laboratory}, 
    city={New York},
    postcode={11973-5000}, 
    country={USA}}
\affiliation[e]{organization={Korea University}, 
    city={Seoul},
    postcode={02841}, 
    country={Korea}}
\affiliation[f]{organization={Institute for Cosmic Ray Research, The University of Tokyo}, 
    city={Kashiwa},
    postcode={277-8582}, 
    state={Chiba},
    country={Japan}}
\affiliation[g]{organization={Shibaura Institute of Technology}, 
    city={Minuma-ku},
    postcode={337-8570}, 
    state={Saitama},
    country={Japan}}
\affiliation[h]{organization={Argonne National Laboratory}, 
    city={Lemont},
    postcode={IL 60439}, 
    country={USA}}       
\affiliation[i]{organization={Sejong University}, 
    city={Seoul},
    postcode={05006}, 
    country={Korea}}
\affiliation[j]{organization={Brookhaven National Laboratory}, 
    city={Upton},
    postcode={11973-5000}, 
    state={New York},
    country={USA}} 
\affiliation[k]{organization={Institute for Space-Earth Environmental Research, Nagoya University}, 
    city={Nagoya},
    postcode={464-8602}, 
    state={Aichi},
    country={Japan}}
\affiliation[l]{organization={Kobayashi-Maskawa Institute for the Origin of Particles and the Universe, Nagoya University}, 
    city={Nagoya},
    postcode={464-8602}, 
    state={Aichi},
    country={Japan}}
\affiliation[m]{organization={Tokushima University}, 
    city={Tokushima},
    postcode={770-8051}, 
    state={Tokushima},   
    country={Japan}}
\affiliation[n]{organization={Advanced Science Research Center, Japan Atomic Energy Agency}, 
    city={Tokai-mura},
    postcode={319-1195}, 
    state={Ibaraki},      
    country={Japan}}
\affiliation[o]{organization={RISE, Waseda University}, 
    city={Shinjuku},
    postcode={162-0044}, 
    state={Tokyo}, 
    country={Japan}}
\affiliation[p]{organization={Department of Physics and Astronomy, University of Catania}, 
    city={Catania},
    postcode={I-95123}, 
    country={Italy}}
\affiliation[q]{organization={INFN Section of Catania}, 
    city={Catania},
    postcode={I-95123}, 
    country={Italy}}
\affiliation[r]{organization={CSFNSM}, 
    city={Catania},
    postcode={I-95123}, 
    country={Italy}}

\begin{abstract}
This letter reports the differential production cross-section of photons in six pseudorapidity regions covering $\eta \, > \, 6.1$,
measured by the RHICf experiment with \pp collisions at $\sqrt{s}$ = 510 GeV conducted in June 2017. In addition, the cross-sections in the three regions of the \xF-\pT~phase space coverage that are the same as those of the LHCf results at $\sqrt{s}$ = 7 and 13 TeV were obtained and compared. Considering the uncertainties, the results were observed to be consistent with both the Feynman scaling law and the model predictions of \EPOSLHC, \QGSJET, \SIBYLL, and \DPMJET, although certain models exhibited weak collision energy dependencies. 
\end{abstract}




\end{frontmatter}


\section{Introduction}
\label{sec:introduction}

A precise understanding of hadronic interactions over a wide energy range is essential for solving the problem of ultra-high-energy cosmic rays (UHECRs). The Pierre Auger Observatory and Telescope Array observe UHECRs with energies above $10^{18}$ eV using the extensive air shower technique, and they have reported their results with high-statistics data~\cite{PAO_spectrum, PAO_Composition_2014_1, TA_Composition_2018}. However, their conclusions on the chemical composition, which is an important parameter for studying the acceleration mechanism of UHECRs at the sources, strongly depend on the choice of the hadronic interaction model used in their air shower simulation. 
The uncertainty of the models is primarily due to the lack of calibration data from accelerator-based experiments. 
Particles produced in the very forward region of interactions carry most of the projectile particle energy, and are crucial to air shower development. The LHC forward~(LHCf) experiment measured the production cross-sections of very forward photons, \pizero, and neutrons at the LHC~\cite{LHCf_Photon_0.9TeVpp, LHCf_Photon_7TeVpp, LHCf_Photon_13TeVpp, LHCf_Pi0_Run1, LHCf_Neutron_7TeVpp, LHCf_13TeVpp_Neutron2}. The maximum center-of-mass collision energy in the previous LHCf measurements is 13 TeV, which corresponds to the energy of $10^{17}$ eV as cosmic ray interactions. 
In addition to this measurement, experimental data at lower collision energies are necessary because an air shower is caused by complex interactions with various interaction energies ranging from GeV to the UHECR energy. 

The RHICf forward~(RHICf) experiment performed a measurement at the Relativistic Heavy Ion Collider~(RHIC), which was similar to that of the LHCf. The aim was to investigate the collision-energy scaling law of forward particle production proposed in Ref.~\cite{FeynmanScaling}, the so-called Feynman scaling, via comparisons with the LHCf results. Whereas LHCf reported the scaling of forward \pizero$\;$ production between $\sqrt{s}$ = 2.76 and 7 TeV~\cite{LHCf_Pi0_Run1}, the scaling can be tested by a LHCf--RHICf comparison over a more than 10 times wider energy range. 
This letter presents the differential production cross-section of photons in six pseudorapidity regions covering $\eta$ $>$ 6.1, as a function of Feynman x (hereafter \xF), which corresponds to the photon energies normialized with the beam energy (255 GeV), based on data acquired with the RHICf detector in June 2017 during $pp$ collisions at $\sqrt{s}$ = 510 GeV.
Furthermore, the results in three additional regions covering the same \xF-\pT~phase space as the LHCf results at $\sqrt{s}$ = 7 and 13 TeV were obtained and subsequently compared to test the Feynman scaling. 

\section{Detector and Data}
\label{sec:rhicf}

\begin{figure}[tb]
\centering
\subfigure[]{
  \centering
  \includegraphics[width=0.4\textwidth]{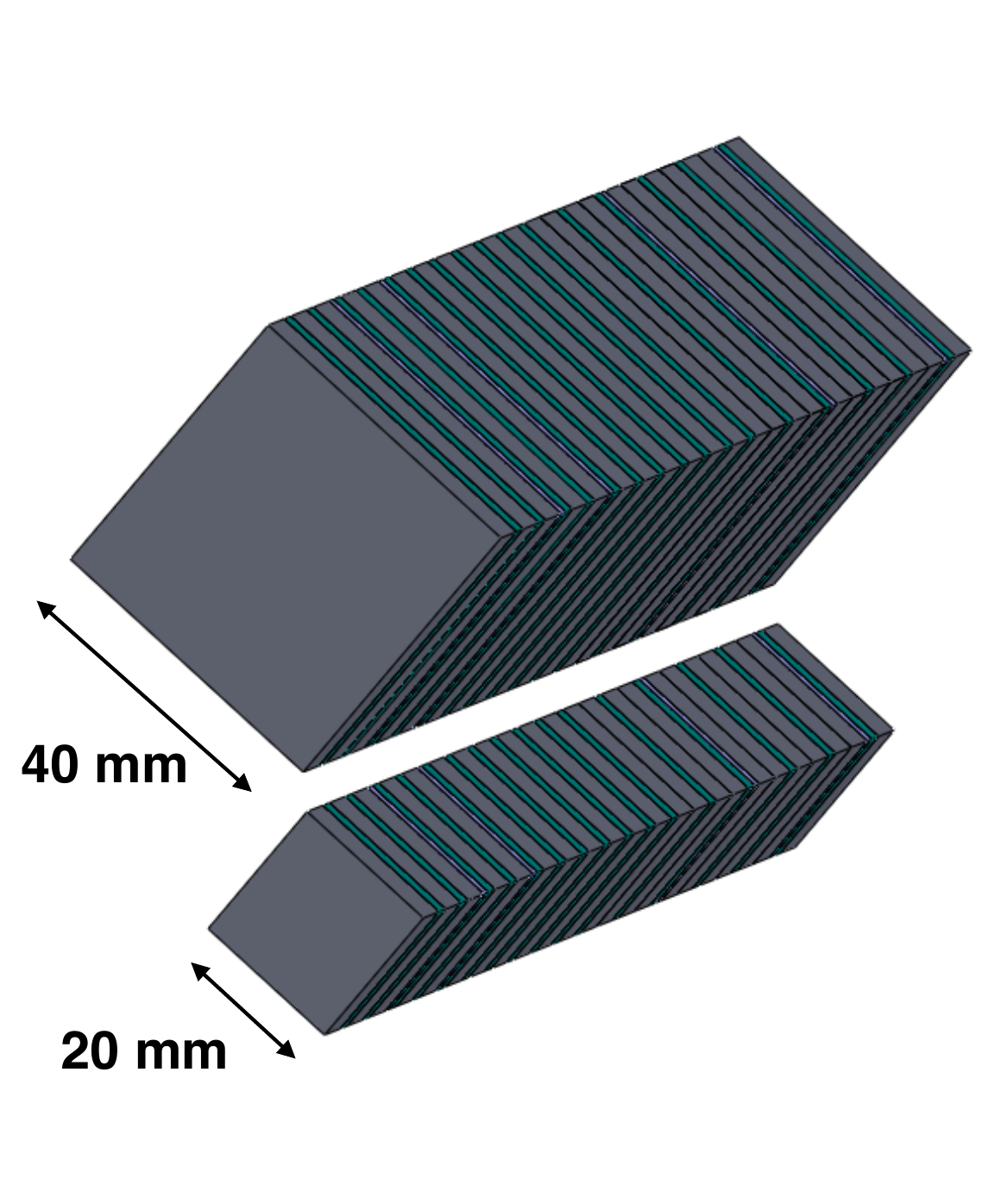}
  }
\subfigure[]{
  \centering
  \includegraphics[width=0.5\textwidth]{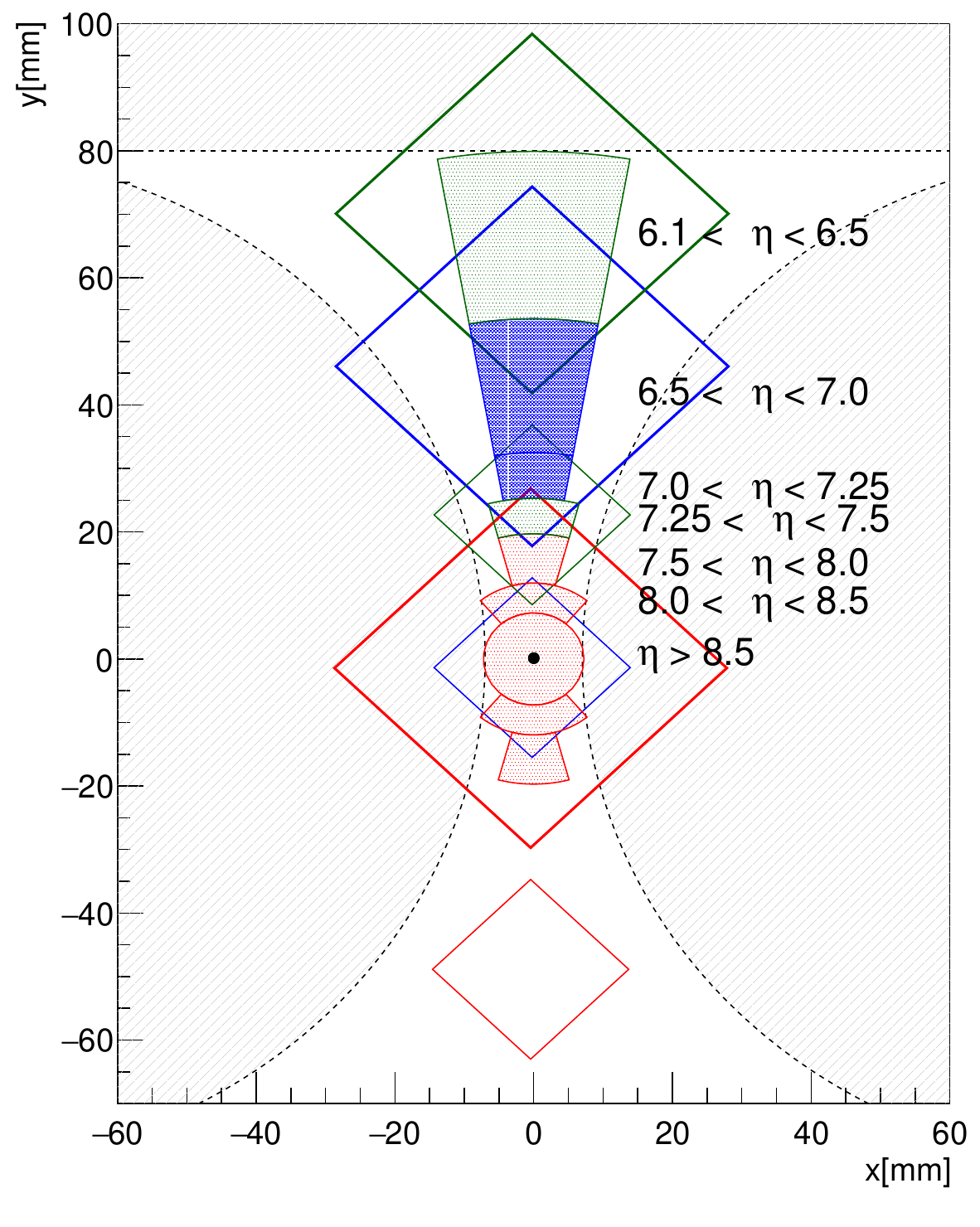}
}
\caption{(a) Schematic view of the RHICf detector. (b) Definitions of analysis regions for the production cross-section measurement viewed from the IP. The origin of the reference frame is centered on the projection of beam center on the detector. The diamond shapes show the calorimeter acceptances at each BOTTOM (red), MIDDLE (blue) and TOP (green) detector position, and the gray shaded areas indicate the regions with thicker beam-line material between the detector and IP. The red, blue and green shaded area show the analysis regions, and each color corresponds to the sample of detector position used for the analysis.  
}
\label{fig:setup}
\end{figure}

The RHICf detector comprises two compact sampling and positioning calorimeters composed of tungsten, 16 $\mathrm{Gd_2SiO_5}$~(GSO) scintillator plates, and 4 XY hodoscopes of GSO bars, as shown in Fig.~\ref{fig:setup}. 
Two calorimeter towers have diamond-shaped acceptances transverse to the beam directions of 20\,mm\,$\times$\,20\,mm and 40\,mm\,$\times$\,40\,mm, which are referred to as the TS and TL, respectively. 
The detector is the former LHCf-Arm1 detector \cite{LHCf_TDR} and its performance has been well studied via beam tests using electron and proton beams at CERN-SPS~\cite{LHCf_Photon_Performance_Makino}. The energy and position resolutions of photons with energies greater than 30 GeV are better than 5\% and 0.2 mm, respectively ~\cite{RHICf_performance}.

The detector was installed 18\,m west of the Solenoidal Tracker at the RHIC (STAR) interaction point~(IP) during the low-luminosity operation period from June 24 to 27, 2017. Because of the magnetic field of the dipole magnet located between the IP and detector, only neutral particles, photons, and neutrons hit the detector. 
Almost all photons hitting the RHICf detector are decay products of $\pi^0$ and $\eta$ mesons, which are produced in $pp$ collisions and immediately decay at the IP. 
The acceptance at the location was restricted because of the shadow of the beam pipe structure located in front of the detector, which corresponds to a pseudorapidity coverage greater than 6.1. The detector was moved vertically to cover the full acceptance, and the data acquisition was performed in three detector positions: BOTTOM~($\Delta y$ = -47.4 mm), MIDDLE~($\Delta y$ = 0 mm), and TOP~($\Delta y$ = +24.04 mm), where $\Delta y$ corresponds to the relative vertical position to MIDDLE~(Fig.~\ref{fig:setup}). 

In this analysis, the data acquired with the two trigger modes, {\it Shower} and {\it High-EM} ~\cite{RHICf_performance} were used. The {\it Shower} trigger was the baseline trigger designed to detect both electromagnetic and hadronic showers induced by photons and neutrons, and a large prescale factor of 8--30 was applied. The trigger efficiency was 100\% for photons with energies greater than 20 GeV. Furthermore, {\it High-EM} triggers were generated only for events with high-energy deposits in the 4th layer of the GSO scintillator to enrich high-energy photon events in the sample. The prescale factor for the {\it High-EM} trigger was approximately 10 times smaller than that for the {\it Shower} trigger. In addition, the trigger efficiencies of the {\it High-EM} trigger were 96\% for photons with energies over 100 GeV. Avoiding the contamination of low-energy photons produced in interactions of secondary particles produced in the primary $pp$ collisions with the beam pipe, only photons with \xF~of greater than 0.1 were analyzed. 

The operation was performed under a special low-luminosity condition. The beam energy was 255 GeV, and the proton beams were radially polarized. The direction of beam polarization was rotated by 90 degrees from vertical polarization around the beam direction. 
The instantaneous luminosity during the operation period was measured at $L$ = $(0.7-1.5) \, \times 10^{31}~\mathrm{cm^{-2}{s^{-1}}}$ using the coincidence rate of signals from the STAR-ZDC detectors~\cite{star_ZDC} located on both the east and west sides of the IP. The collision pileup rate was approximately 0.05 collisions per bunch crossing. Considering a small acceptance of the detector, the event pileup is negligible in this measurement.
The recorded integral luminosities of the Shower (High-EM) trigger sample were 4.5~(27.6), 12.0~(96.0), and 20.2~(120.0) $\mathrm{nb^{-1}}$ for the BOTTOM, MIDDLE, and TOP detector positions, respectively.

For comparison with the experimental data, Monte Carlo~(MC) event samples were generated using the CRMC framework~\cite{CRMC} with four hadronic interaction models: \EPOSLHC~\cite{EPOSLHC}, \QGSJET~\cite{QGSJET}, \SIBYLL~\cite{Sibyll2.3d}, and \DPMJET~\cite{DPMJET}. These models are widely used in simulations of extensive air showers induced by high-energy cosmic rays. All the models were tuned using early LHC data, and are commonly referred to as post-LHC hadronic interaction models.

\section{Analysis}
\label{sec:analysis}

\subsection{Event reconstruction and selection}

For the analysis, the event reconstruction algorithm reported in Ref.~\cite{LHCf_Photon_13TeVpp, RHICf_performance} was employed. 
The criteria in this analysis were reoptimized for the RHICf configuration using a MC simulation of the detector implemented with the GEANT4 library~\cite{GEANT4}.
Events that satisfied the particle identification (PID) criteria for photons and the single-hit condition were selected, corresponding to cases in which only one particle struck a calorimeter tower.

The energy scale of the calorimeters was calibrated using the factor estimated in a study using \pizero$\;$ events where photon pairs were observed in the calorimeters. Thereafter, the invariant mass was calculated from the energies and hit positions of a photon pair, assuming a decay vertex at the IP. The factor was calculated to shift the peak of the mass distribution to the rest mass of the \pizero. 

Analysis regions were defined for the production cross-section measurement through the division of the pseudorapidity coverage at the location from 6.1 to infinity by six; \etaregion{6.1}{6.5}, \etaregion{6.5}{7.0}, \etaregion{7.0}{7.5} \etaregion{7.5}{8.0}, \etaregion{8.0}{8.5}, and \etaregionZero. The region is a sector-shaped area defined by the range of the distance from the beam center\footnote{ The beam center was defined as the projection of the beam direction at the IP on the detector surface, which was measured using data~\cite{RHICf_performance}.} $R$ and the azimuthal angle $\phi$, whereas the region for $\eta$ $>$ 8.5 is a disk shape with $R$ $<$ 7.25 mm~(Fig.~\ref{fig:setup}). 
The $R$ and $\phi$ ranges of each region and the data set~(detector position and calorimeter tower) used for the analysis regions are listed in Tab.~\ref{tab:dataset}. However, region \etaregion{7.0}{7.5} was not covered by a single dataset. Therefore, subregions corresponding to \etaregion{7.00}{7.25} and \etaregion{7.25}{7.50} were defined, which were combined to obtain the final result. 

In addition to these regions, three analysis regions were defined to test the Feynman scaling law. Each region covered the same \xF-\pT$\;$ phase space as one of the LHCf measurements; \pT$\;$$<$ 0.124 \xF$\;$[GeV/c], 0.87 \xF$\;$ [GeV/c] $\;<$ \pT$\;$$<$ 1.04 \xF$\;$ [GeV/c] for $\eta\,>\,10.94$ and $8.81\,<\,\eta\,<\,8.99$ at $\sqrt{s}$ = 7 TeV~\cite{LHCf_Photon_7TeVpp}, respectively, and \pT$\;$$<$ 0.23 \xF$\;$[GeV/c] for $\eta\,>\,10.94$ at $\sqrt{s}$ = 13 TeV~\cite{LHCf_Photon_13TeVpp}. The corresponding pseudorapidity regions at  $\sqrt{s}$ = 510 GeV are $ \eta\,>\,8.32$, $6.19\,<\,\eta\,<\,6.38$, and $ \eta\,>\,7.70$, respectively.

The previous RHICf measurement discovered a large transverse single-spin asymmetry of forward \pizero$\;$production using the same dataset~\cite{RHICf_pi0_An}.
Owing to the \pizero$\;$asymmetry inducing dependency of the photon spectrum on the spin direction of the incoming beams to the detector, the events were categorized with the spin direction {\it inward} or {\it outward} of the bunch associated with each event. The difference of the integral luminosity between the spin samples was only 1--4\%.

\begin{table}[hbtp]
  \caption{ Analysis regions and data sets used for this analysis. }
  \label{tab:dataset}
  \centering
  \begin{tabular}{lccc}
    \hline
    Region  & R [mm] & $\phi$ [degrees]  &  Dataset  \\
    \hline \hline
    \etaregionZero & $<$ 7.3 & 0 - 360  & BOTTOM-TL \\
    \etaregion{8.0}{8.5} & 7.3 - 12.0 & 50 - 130, 230 - 310  & BOTTOM-TL \\
    \etaregion{7.5}{8.0} & 12.0 - 19.7 & 75 - 105, 255 - 285  & BOTTOM-TL \\
    \etaregion{7.0}{7.5}  &   &   \\
    \quad \etaregion{7.25}{7.50} & 19.7 - 25.3 &  75 - 105 &  TOP-TS \\
    \quad \etaregion{7.00}{7.25} & 25.3 - 32.5 &  80 - 100 &  MIDDLE-TL \\
    \etaregion{6.5}{7.0} & 32.5 - 53.6 & 80 - 100  & MIDDLE-TL \\   
    \etaregion{6.1}{6.5} & 53.6 - 79.9 & 80 - 100  & TOP-TL \\    
    \hline 
    \pT~$<$~0.124 \xF & $<$ 8.7 & 45 - 135, 225 - 315 & MIDDLE-TL \\ 
    \pT~$<$~0.23 \xF & $<$ 16.1 & 70 - 110, 250 - 290 & MIDDLE-TL \\ 
    0.87 \xF $<$ \pT $<$ 1.04 \xF & 60.7 - 72.8 & 80 - 100 & TOP-TL \\ 
    \hline
  \end{tabular}
\end{table}

\subsection{Corrections}
\label{sec:correction}

The final result of the differential production cross-section of photons, $d\,\sigma_{\gamma}/ d\,x_{\mathrm{F}}$, was obtained considering the number of events passing the event selection $N^{\mathrm{single-\gamma}}_{j}$ in the {\it j}-th spin sample ({\it j} = {\it inward} or {\it outward}) with a bin-by-bin correction procedure: 
\begin{equation}
  \frac{d\,\sigma_{\gamma}}{d\,x_{\mathrm{F}}} =  C^{\mathrm{geo}}\, C^{c\tau}\, C^{\mathrm{MC}} \,\sum_{j} \frac {C^{\mathrm{PID}}_{j}\, ( 1 - R^{\mathrm{BKG}}_{j} ) } {2 \, {L_{j}}\, } \frac{ \Delta N^{\mathrm{single-\gamma}}_{j}}{\Delta x_{\mathrm{F}}},
\end{equation}
where $C^{\mathrm{geo}}$, $C^{c\tau}$, $C^{\mathrm{MC}}$, $C^{\mathrm{PID}}_{j}$, and $R^{\mathrm{BKG}}_{j}$ are correction factors and a background fraction described below. $C^{c\tau}$, $C^{\mathrm{MC}}$, and $C^{\mathrm{PID}}_{j}$ are functions of $\eta$ and \xF, while $C^{\mathrm{geo}}$ and $R^{\mathrm{BKG}}_{j}$ depend only on $\eta$. 
$L_{j}$ is the recorded integral luminosity of the beams in the {\it j}-th spin direction. Subsequently, the PID purity correction $C^{\mathrm{PID}}_{j}$ and the beam-gas background $R^{\mathrm{BKG}}_{j}$ were estimated for each spin sample, whereas the other correction factors were common. 
The difference of photon cross-sections between the two spin samples was 20\% at maximum, which originated from the \pizero$\;$asymmetry. Consequently, upon combining the two spin samples, the effect of spin asymmetry was expected to be negligible in the final result even considering the systematic uncertainty of the integral luminosity. 

\begin{itemize}
\item Beam-gas background $R^{\mathrm{BKG}}_{j}$; \\
The background photons produced in collisions between the beam protons and residual gas in the beam pipe were estimated using the events associated with the non-colliding bunches of the incoming beam to the detector. Further, the background fraction $R^{\mathrm{BKG}}_{j}$ was calculated for each spin direction sample by considering the difference in the total beam intensities between the colliding and non-colliding bunches~\cite{RHICf_performance}. The estimated background factors were 1.5 -- 2.5\%, which was dependent on the beam fill. The factors were consistent with a constant in \xF, however, possible energy dependency was considered as a systematic uncertainty.  
\item PID purity correction $C^{\mathrm{PID}}_{j}$; \\ 
In this analysis, particle identification was performed by introducing a PID estimator $L_{90\%}$, which was defined as the longitudinal depth at which the energy deposit integral measured by the sampling layers reached 90\% of the total energy deposition. Subsequently, the purity of the photons in the selected events for each energy bin was estimated from the template-fit results of the $L_{90\%}$ distribution with MC distributions for photons and hadrons~\cite{RHICf_performance}. The correction factor $C^{\mathrm{PID}}_{j}$~(=~1/purity) was typically 0.9 -- 1.0. 
\item MC based correction factor $C^{\mathrm{MC}}$; \\
This factor was introduced as an overall correction of several contributions related to event reconstruction: inefficiencies of the trigger, PID selection and single-hit selection, misreconstruction of multi-hit events (two particles hit in a tower) as single-hit events, and recovery of the photon yield in the multi-hit events rejected by the single-hit selection. 
The total of these contributions was estimated using the detector simulation with $pp$ event generation based on the \EPOSLHC~or \QGSJET~interaction models, which reproduced the photon results of the LHCf experiment~\cite{LHCf_Photon_13TeVpp}. The correction factor is defined as
\begin{equation}
C^{\mathrm{MC}} = N^{\mathrm{\gamma}}_{\mathrm{MC,true}}/N^{\mathrm{single-\gamma}}_{\mathrm{MC,rec}},
\end{equation}
where $N^{\mathrm{\gamma}}_{\mathrm{MC,true}}$ is the number of photons hitting the analysis region, and $N^{\mathrm{single-\gamma}}_{\mathrm{MC,rec}}$ is the number of events reconstructed as single photons with the reconstructed hit position in the region. 
It is noted that hadron contamination was not considered in $N^{\mathrm{single-\gamma}}_{\mathrm{MC,rec}}$ because it should be corrected using the PID correction. Thus, the average of the results obtained using the two interaction models was used as the correction factor. The typical value of the factor was 1.1 -- 1.2. The difference from unity mainly originates from the inefficiency of the PID selection ($\sim0.1$) and the recovery of the photon yield in multi-hit events (0.02 -- 0.08). Moreover, the interaction-model dependency of the factor is primarily caused by the difference in multi-hit events, and it was assigned as a systematic uncertainty.
\item Correction of long-lived particle contribution $C^{c\tau}$; \\
The dominant sources of photons measured by the detector are the \pizero~and $\eta$ decays produced at collisions, while decays of long-lived particles such as $K^0$ and $\Lambda$ along the beam pipe located between the IP and the detector contribute to the photon yield. In this analysis, the measured photons are defined as photons either directly produced in $pp$ interactions or from subsequent decays of directly produced particles with mean lifetime  ($c\tau$) smaller than 1 cm. 
The contribution of long-lived particles with $c\tau\,>\,1\,\mathrm{cm}$ was estimated using an MC simulation by comparing the photon flux at the IP with that after transportation from the IP to the detector location. It was removed via the application of the correction factor. 
The contribution was relatively large in the low \xF$\;$regions, typically 8\% of the total yield at \xF$\;$ approximately 0.1, whereas it was less than 1\% in the high \xF$\;$region of more than 0.5. The correction factors were calculated as the average of the results obtained using the four interaction models: \EPOSLHC$\;$, \QGSJET$\;$, \SIBYLL, and \DPMJET. Furthermore, the systematic uncertainty of the factor was estimated as the maximum deviation from the average of each model result. 
\item Geometrical correction factor $C^{\mathrm{geo}}$; \\
A correction factor was introduced to correct for the limitated $\phi$ range of the detector, and it was calculated as $C^{\mathrm{geo}} = 360^\circ / \Delta \phi$. 
\end{itemize}

\subsection{Systematic uncertainties}
\label{sec:systematic}

The following contributions were considered as systematic uncertainties of the measured production cross-section, in addition to those related to the correction factors $R^{\mathrm{BKG}}$, $C^{\mathrm{MC}}$, and $C^{c\tau}$ described in Sec.~\ref{sec:correction}.

\begin{itemize}
\item Energy scale \\
Three components were considered as the systematic uncertainty of the energy scale. The first was the stability of the energy scale during the operation, which was monitored using the peak position of the \pizero$\;$ mass on the reconstructed mass distribution, and it was stable within $\pm1\%$~\cite{RHICf_performance}. The second was the nonlinearity. It was observed that the peak mass value of \pizero$\;$ was shifted with increasing \pizero$\;$ energy corresponding to the total energy of the photon pair ~\cite{RHICf_performance}. Assuming that the shift originates only from the nonlinearity of the energy scale conservatively, the corrected photon energy $E'$ was obtained as $E' = (1.03 - 0.025\,(E/100 \mathrm{GeV})) E$, where $E$ is the original value of the reconstructed energy. 
The last factor was the nonuniformity of the energy scale. Although beam-test measurements showed that the calorimeter response was uniform within approximately 1\%~\cite{LHCf_Photon_Performance_Makino}, a more conservative uncertainty was estimated from the consistency of the spectra obtained by dividing the $\phi$ range in the data.
The impact of each energy-scale uncertainty on the final result was estimated by repeating the analysis by artificially changing the energy scale. Owing to the steep slope of the \xF\; spectrum, large systematic uncertainties of over 50\% resulted in the highest energy bin.
\item PID selection \\
The experimental distribution of the PID estimator $L_{90\%}$ was not perfectly reproduced by the template distributions obtained from the MC simulations. The systematic effect of the difference on the final result through the purity and efficiency estimation of PID was estimated using a method similar to that used in Ref.~\cite{LHCf_Photon_7TeVpp, LHCf_Neutron_7TeVpp}; Another template-fitting method that allowed artificial displacement and widening of the template distributions was performed. Consequently, the difference in the results between the methods was assigned as systematic uncertainty. 
\item Beam center \\
The effect of the uncertainty of the beam-center determination was evaluated via repeating the analysis procedure by shifting the beam center by $\pm$1 mm vertically or horizontally. The maximum deviation in these results from the original results was assigned as the systematic uncertainty. 
\item Luminosity \\
The uncertainty of the integral luminosity is $\pm5\%$, which is due to the uncertainty of the conversion factor from the ZDC signal rate to the absolute luminosity determined from the Vernier scans using the method similar to that described in Ref.~\cite{Luminosity}.
\end{itemize}

\begin{figure}[tb]
    \centering
    \includegraphics[width=0.5 \textwidth]{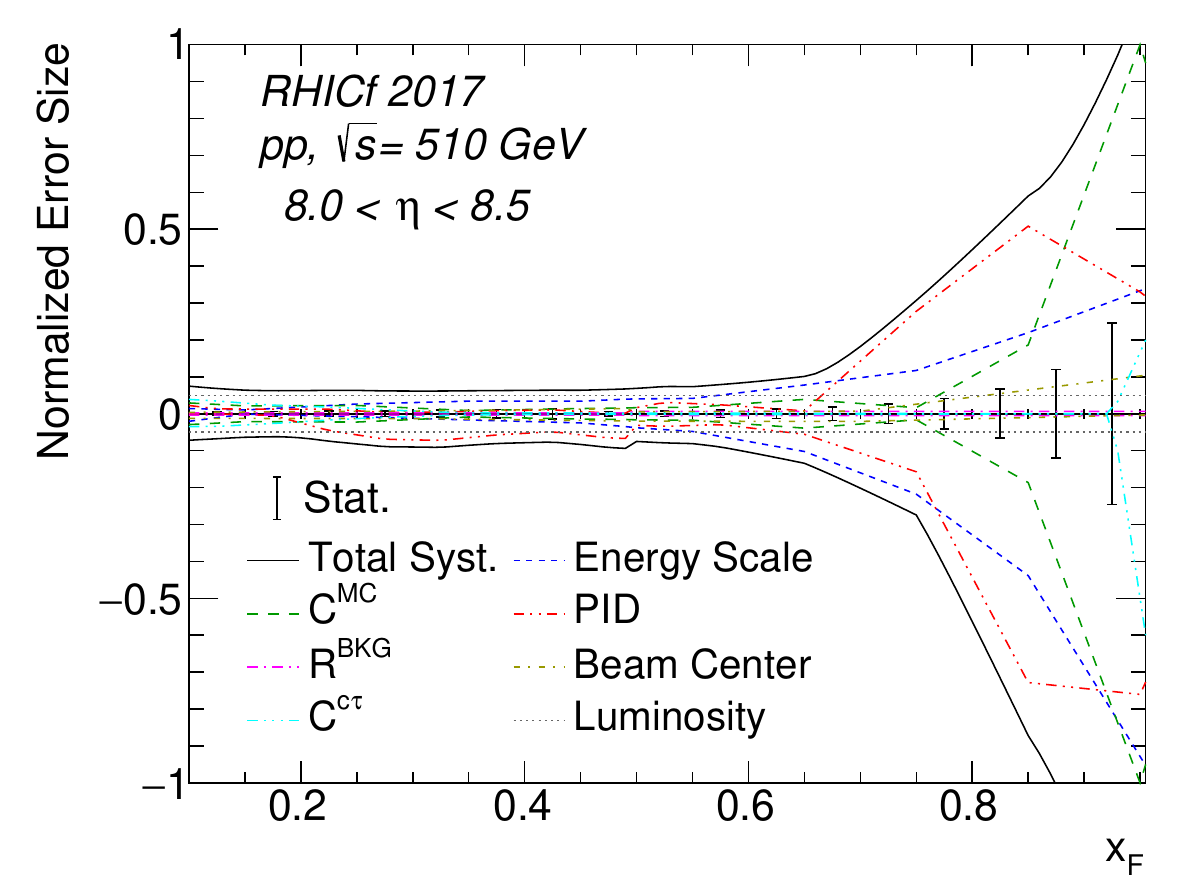}
    \caption{Systematic and statistical uncertainties of the photon production cross-section measurement for the pseudorapidity region \etaregion{8.0}{8.5}. The coloured and dashed lines indicate the estimated systematic uncertainties after normalization based on the mean values of the experimental data. The black line indicates the total systematic uncertainties, calculated as the quadratic summations of all uncertainties. The error bars indicates the size of statistical uncertainties.}
    \label{fig:syserr}
\end{figure}

The total systematic uncertainty was calculated as the quadratic summations of these uncertainties. 
Figure~\ref{fig:syserr} shows the estimated systematic and statistical uncertainties for the pseudorapidity region \etaregion{8.0}{8.5} as a typical example. The systematic uncertainties dominate the uncertainties in any \xF\; and pseudorapidity bins. 

\section{Results}
\label{sec:results}

\subsection{Differential production cross-section}

\begin{figure}[tb]
    \centering
    \includegraphics[width=1.0 \textwidth]{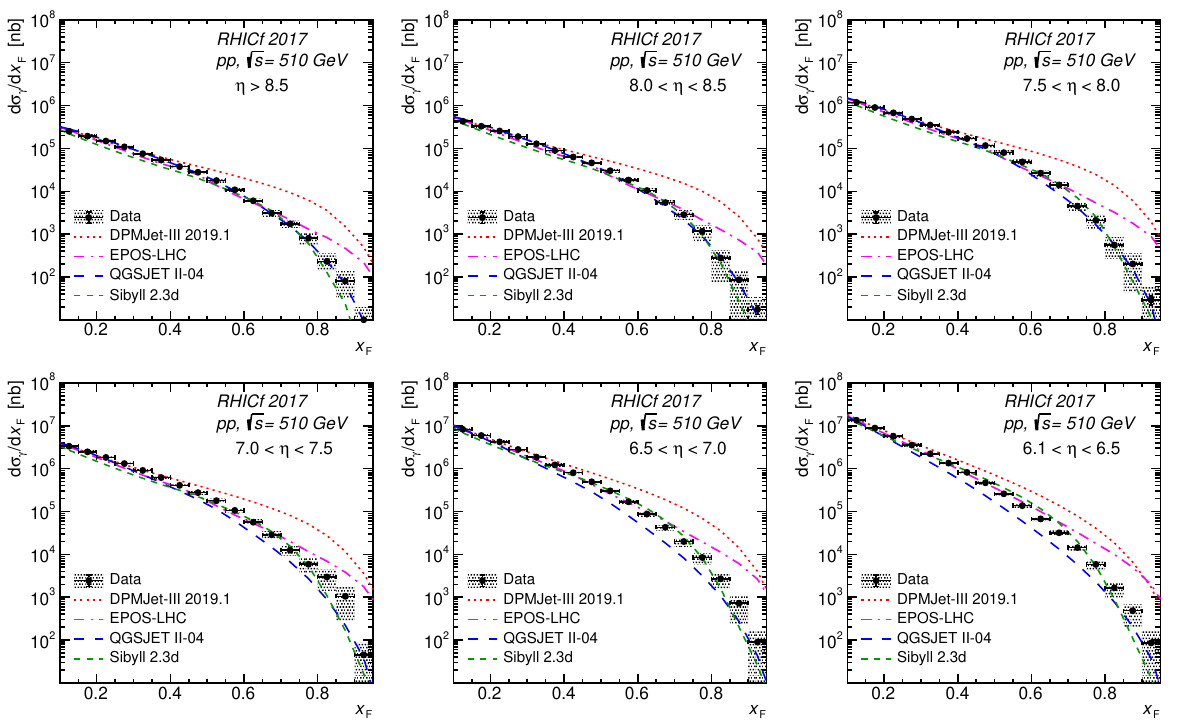}
    \caption{Differential photon production cross-section measured by the RHICf detector as a function of \xF.  Each figure represents that for one of the pseudorapidity regions: $\eta$ $>$ 8.5, 8.0 $<$ $\eta$ $<$ 8.5, 7.5 $<$ $\eta$ $<$ 8.0, 7.0 $<$ $\eta$ $<$ 7.5, 6.5 $<$ $\eta$ $<$ 7.0, and 6.1 $<$ $\eta$ $<$ 6.5. The bars and hatched areas correspond to the statistical and systematic uncertainties, respectively. Coloured lines indicate the MC predictions of \DPMJET, \EPOSLHC, \QGSJET, and \SIBYLL.}
    \label{fig:spectra}
\end{figure}

\begin{figure}[tb]
    \centering
    \includegraphics[width=1.0 \textwidth]{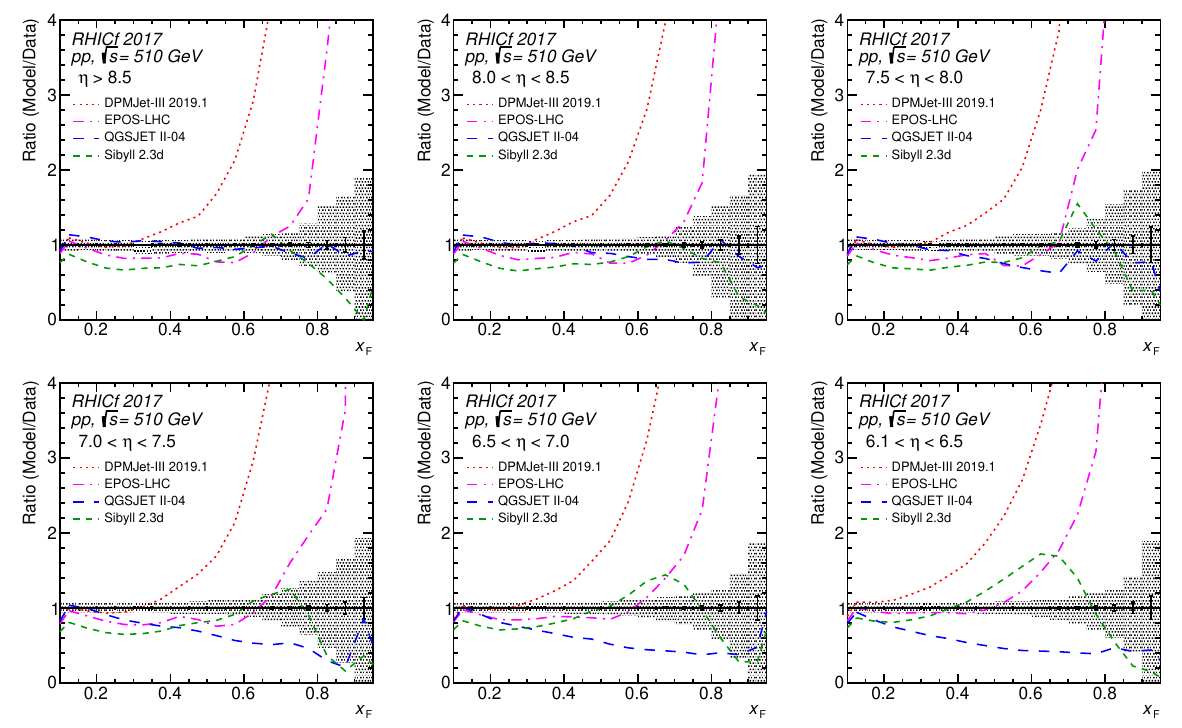}
    \caption{Ratio of differential photon production cross-sections predicted by hadronic interaction models to the experimental result. The bars and hatched areas around one correspond to the normalized statistical and systematic uncertainties of the data, respectively. }.
    \label{fig:spectra_ratio}
\end{figure}

Figure~\ref{fig:spectra} presents the differential production cross-section of photons as a function of \xF, $d\,\sigma_{\gamma}/d\,x_{\mathrm{F}}$, for the six pseudorapidity intervals measured by the RHICf detector. The data points in the \xF$\;$range below 0.45 were obtained from the Shower trigger samples, while those in the higher \xF$\;$region were obtained from the High-EM trigger samples because of larger statistics. The coloured lines in Fig.~\ref{fig:spectra} represent the prediction of the interaction models, \EPOSLHC, \QGSJET, \SIBYLL, and \DPMJET. Figure~\ref{fig:spectra_ratio} presents the ratios of these predictions to the experimental results.
\EPOSLHC$\;$and \DPMJET$\;$ are consistent with the experimental result in the low \xF$\;$region below 0.6 and 0.3, respectively, for any pseudorapidity regions, although they predict a larger flux than the data in the higher \xF$\;$region. \QGSJET$\;$and \SIBYLL$\;$well reproduce the shape of the \xF$\;$spectra in the highest pseudorapidity regions; however, the slope of their spectra becomes progressively softer and harder, respectively, than that of the data in the lower pseudorapidity regions. This can be interpreted as the difference in \pT$\;$distributions between the data and the model. Thus, these features of the data-model comparisons are very similar to those at the LHC energies reported by the LHCf~\cite{LHCf_Photon_0.9TeVpp, LHCf_Photon_7TeVpp,LHCf_Photon_13TeVpp}, thereby suggesting that the source of the difference is common to a wide range of collision energies.

\begin{figure}[tb]
    \centering
    \includegraphics[width=0.4\textwidth]{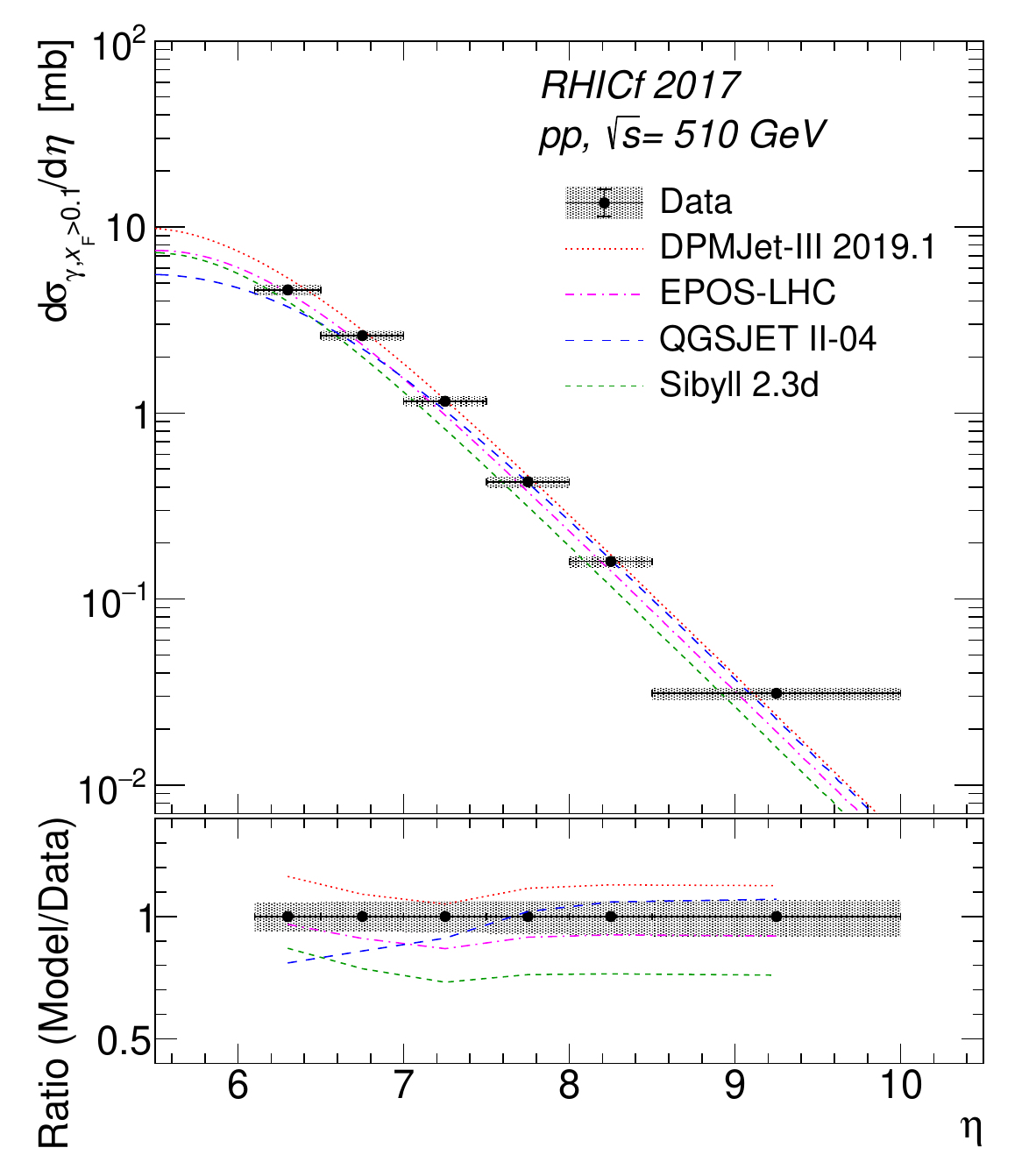}
     \includegraphics[width=0.4\textwidth]{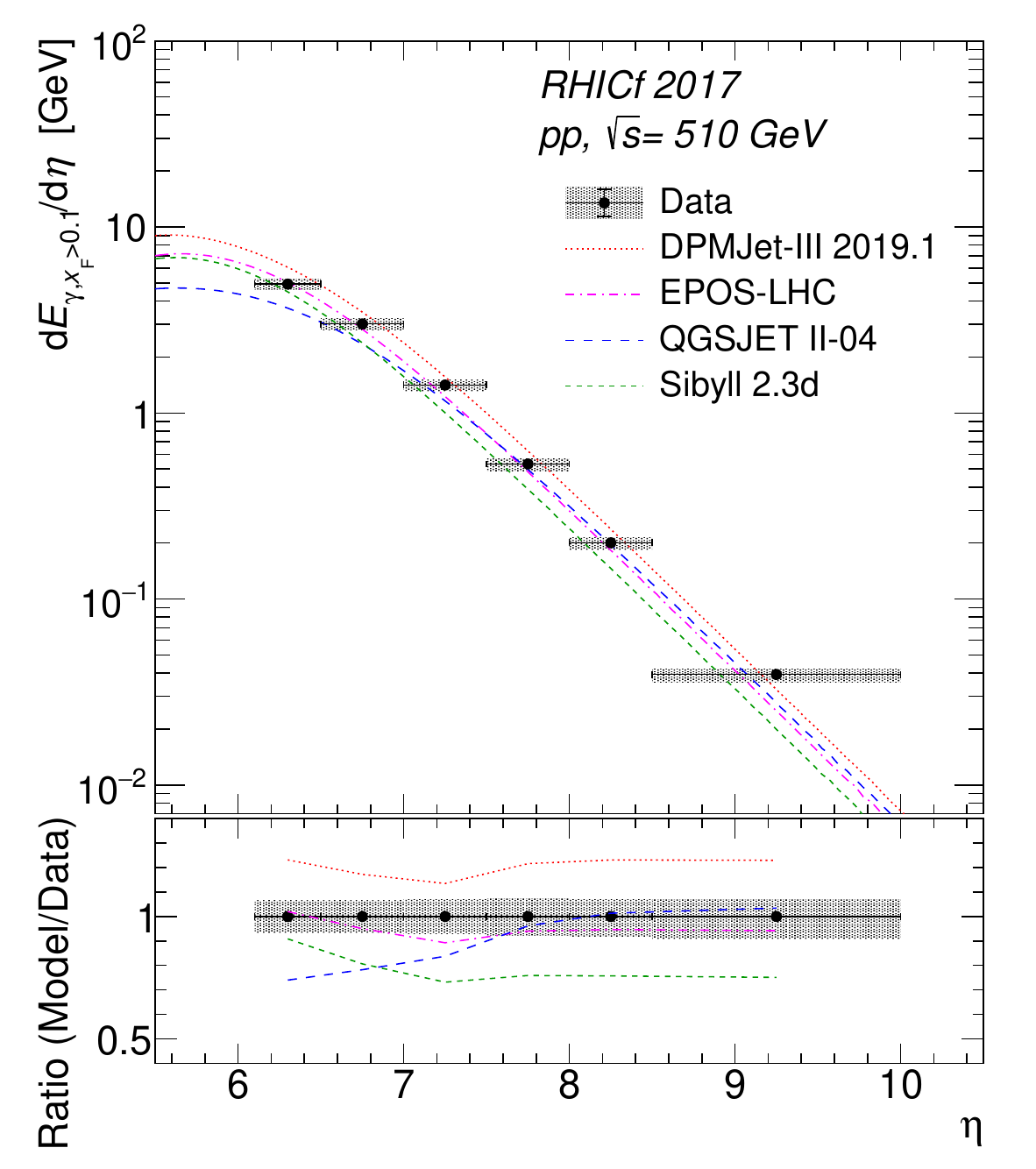}
    \caption{Differential cross-section $d\,\sigma_{\gamma,x_{\mathrm{F}}>0.1}/d\,\eta$ (left) and differential energy flow $d\,E_{\gamma,x_{\mathrm{F}}>0.1}/d\,\eta$ of photons with $x_{\mathrm{F}}\,>\,0.1$ produced in $pp$ collisions at $\sqrt{s}$ = 510 GeV. The bars and hatched areas represent the statistical and systematic uncertainties of the experimental data, respectively. Coloured lines indicates the MC predictions. The lower panels show the ratios of the MC predictions to the experimental data.
    }
    \label{fig:integral}
\end{figure}

The differential photon production cross-section $d\,\sigma_{\gamma,x_{\mathrm{F}}>0.1}/d\,\eta$ as a function of $\eta$ for \xF~ $>$ 0.1 is shown in Fig.~\ref{fig:integral} (left). 
It was calculated from the integration of the obtained differential cross section $d\sigma_{\gamma}/dx_{\mathrm{F}}$, 
\begin{equation}
     \frac{d\,\sigma_{\gamma,x_{\mathrm{F}}>0.1}}{d\,\eta} =  \frac{1}{\Delta \eta} \sum_{i}  \left. \frac{d\,\sigma_{\gamma}}{d\,x_{\mathrm{F}}} \right|_{i} \Delta\, x_{\mathrm{F},i} 
     \label{eq-integralsigma}
\end{equation}
where $\Delta\eta$ is the width of the pseudorapidity range, and $\Delta x_{\mathrm{F},i}$ is the width of the $i$-th \xF~bin in Fig.~\ref{fig:spectra}.  The value of the largest pseudorapidity bin ($\eta$ $>$ 8.5) was computed with replacement of the pseudorapidity region $\eta$ $>$ 8.5 to 8.5 $<$ $\eta$ $<$ 10.0 due to technical reasons. 

The differential energy flow $dE_{\gamma,x_{\mathrm{F}}>0.1}/d\eta$ for \xF~$>$ 0.1 was shown in Fig.~\ref{fig:integral} (right), which was obtained by
\begin{equation}
        \frac{d\,E_{\gamma,x_{\mathrm{F}}>0.1}}{d\,\eta} = \frac{E_{\mathrm{beam}}}{\sigma_{\mathrm{inela}}\, \Delta \eta} \,\sum_{i} \left. \frac{d\,\sigma_{\gamma}}{d\,x_{\mathrm{F}}} \right|_{i} \, \langle x_{\mathrm{F}} \rangle_{i}\, \Delta \,x_{\mathrm{F},i}\; 
        \label{eq-integralflux},
\end{equation}
where $E_{\mathrm{beam}}$ is the proton beam energy of 255 GeV, $\sigma_{\mathrm{inela}}$ is the inelastic cross section, and $\langle x_{\mathrm{F}} \rangle_{i}$ is the average of \xF~in the $i$-th \xF~bin. 
The $pp$ inelastic cross section at $\sqrt{s}$ = 510 GeV was estimated to be $(48.3\,\pm\,1.1)$ mb using the fit results of total and elastic cross-sections by the COMPETE~\cite{COMPETEfit} and TOTEM~\cite{TOTEM_13TeV} collaborations, respectively. 
The difference between $\langle x_{\mathrm{F}} \rangle$ and the center value of the \xF~bin was less than 1.5\% in any \xF~bins. 
The values of lower \xF~bins contribute more to the integration. Therefore, the contribution of photons with \xF~$<$ 0.1, which were outside the measurement acceptance, is not negligible in the total energy flow. No correction was applied for this unmeasured region. Instead, its contribution was estimated using \DPMJET, which reproduces the data well in the low \xF~region, and was found to be approximately 12--20\% of the total energy flow.  The estimated contribution for each $\eta$ bin is listed in Tab.~\ref{tab:result3}. 

\subsection{Collision-energy scaling}

\begin{figure}[tb]
    \centering
    \includegraphics[width=1.0 \textwidth]{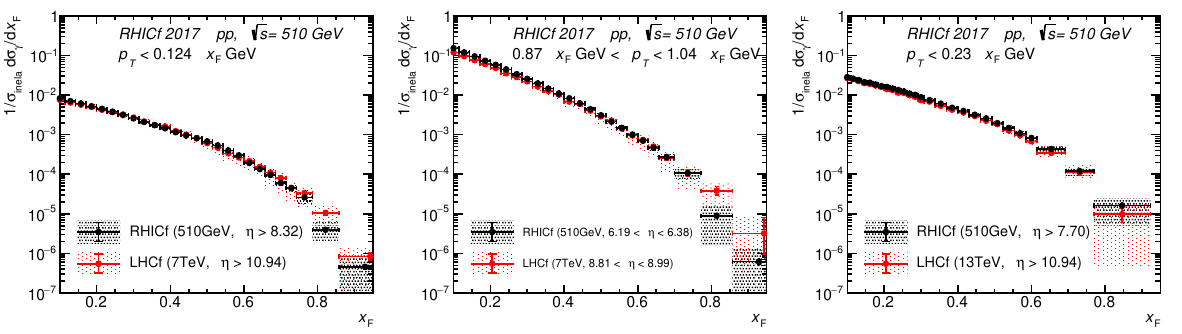}
    \caption{Comparison of the measured cross-section with the results of LHCf at $\sqrt{s}$ = 7 and 13 TeV~\cite{LHCf_Photon_7TeVpp,LHCf_Photon_13TeVpp}. The two left panels show the comparison with the LHCf result at $\sqrt{s}$ = 7 TeV and the pseudorapidity region of $\eta\,>\,10.94$ (left) and $8.81\,<\,\eta\,<\,8.99$ (middle). The right panel shows a comparison between $\sqrt{s}$ = 13 TeV and $\eta\,>\,10.94$. The bars and hatched areas correspond to statistical uncertainties and quadratic summation of statistics and systematic uncertainties, respectively.}.
    \label{fig:escalespe}
\end{figure}

\begin{figure}[tb]
    \centering
    \includegraphics[width=1.0 \textwidth]{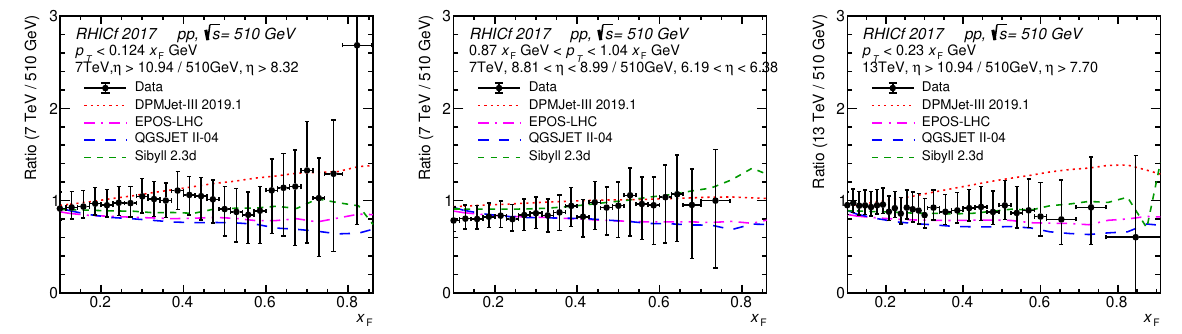}
    \caption{Ratios of the RHICf results to the LHCf results~\cite{LHCf_Photon_7TeVpp,LHCf_Photon_13TeVpp}. The error bars represent the uncertainties calculated as a quadratic summation of the uncertainties in these results. Coloured lines indicate MC predictions.}.
    \label{fig:escale}
\end{figure}

The measured cross-sections were compared with the results of LHCf at $pp$ collisions at $\sqrt{s}$ = 7 and 13 TeV to test the collision energy dependency of forward photon production.
Fig.~\ref{fig:escalespe} shows the results obtained for the same coverage of \xF-\pT$\;$phase space as that of the LHCf results.
In this comparison, an additional normalization factor $1/\sigma_{\mathrm{inela}}$ was introduced. For normalization of the LHCf results, the values of $\sigma_{\mathrm{inela}}$ = $(72.9\,\pm\,1.5)$ mb for $\sqrt{s}$ = 7 TeV and $(79.5\,\pm\,1.8)$ mb for $\sqrt{s}$ = 13 TeV reported in Ref.~\cite{TOTEM_7TeV, TOTEM_13TeV} were used\footnote{A slightly lower value $\sigma_{\mathrm{inela}}$ = 71.5 mb for $\sqrt{s}$ = 7 TeV was used in Ref.~\cite{LHCf_Photon_7TeVpp}. Their results were rescaled using this ratio.}. 
The RHICf result was consistent with the LHCf results within the uncertainties. The ratios shown in Fig.~\ref{fig:escale} are consistent with unity within the uncertainties in all the measured points, while the centre values are slightly low in the small \xF~region of the left and middle panels. This feature can be interpreted as the difference in analysis methods: corrections of photon yield in the rejected multi-hit events and photons from long-lived particle decays were not performed in the LHCf 7 TeV results, whereas they were performed in both the RHICf and LHCf 13 TeV results. Considering the difference, these ratios may be rather close to unity. 

All four models reproduce the ratio obtained, as shown in Fig.~\ref{fig:escale}. Certain models predict a weak \xF$\;$dependency of the ratio; however, the dependency could not be confirmed owing to the large uncertainties in the results. 

\afterpage{\clearpage}

\section{Summary}
\label{sec:summary}

This study measured the differential production cross-section of photons in the very forward region $\eta\,>\,6.1$ with $pp$ collisions at $\sqrt{s}$ = 510 GeV. 
Through comparisons of the obtained results with those from the LHCf experiment, the Feynman scaling law of forward photon production in a wide energy range from 510 GeV to 13 TeV was confirmed within the uncertainties. However, the precision is not sufficient to probe the weak \xF$\;$dependency predicted by certain models, and future studies must focus on reducing the uncertainties together with the LHCf collaboration. Moreover, tests of the Feynman scaling law with other neutral hadrons such as \pizero s and neutrons are an interesting concept and will be addressed in future publications. They will contribute to improving the predictions of models with intermediate and even higher collision energies than LHC.

\section*{Acknowledgements}

We thank the staff at BNL, STAR Collaboration, and PHENIX Collaboration for supporting the experiments. Further, we acknowledge the essential contributions of STAR members toward the successful operation of RHICf. 
We are also grateful to BNL Collider-Accelerator Department for smooth commissioning and operation of beams, and providing beam parameters used in this analysis such as luminosity. 
This work was partly supported by the U.S.--Japan Science and Technology Cooperation Program in High-Energy Physics, JSPS KAKENHI (No. JP26247037 and No. JP18H01227), the joint research program of the Institute for Cosmic Ray Research (ICRR), University of Tokyo, and the National Research Foundation of Korea (No. 2018R1A5A1025563), and “UNICT 2020-22 Linea 2” program, University of Catania.



\afterpage{\clearpage}

\appendix

\section{Cross-section tables}
\label{sec:app}

\begin{table}
  \caption{ Differential photon production cross-section $d\,\sigma_{\gamma}/d\,x_{\mathrm{F}}$ [nb] for each \xF~bin in the pseudorapidity ranges: \etaregionZero, \etaregion{8.0}{8.5} and \etaregion{7.5}{8.0}. Upper and lower total uncertainties are also reported. }
  \label{tab:result1}
  \centering
  \begin{tabular}{cccc}
    \hline
     & \multicolumn{3}{c}{ $d\,\sigma_{\gamma}/d\,x_{\mathrm{F}}$ [nb]} \\ \cline{2-4}
    \xF  & \etaregionZero & \etaregion{8.0}{8.5}  &  \etaregion{7.5}{8.0}  \\
    \hline 
0.10 - 0.15 & $(2.51\,^{+0.18}_{-0.18}) \times 10^{5}$ & $(4.30\,^{+0.30}_{-0.29}) \times 10^{5}$ & $(1.18\,^{+0.08}_{-0.08}) \times 10^{6}$  \\
0.15 - 0.20 & $(1.93\,^{+0.13}_{-0.13}) \times 10^{5}$ & $(3.29\,^{+0.21}_{-0.21}) \times 10^{5}$ & $(9.00\,^{+0.56}_{-0.56}) \times 10^{5}$  \\
0.20 - 0.25 & $(1.49\,^{+0.10}_{-0.11}) \times 10^{5}$ & $(2.55\,^{+0.16}_{-0.19}) \times 10^{5}$ & $(6.76\,^{+0.42}_{-0.44}) \times 10^{5}$  \\
0.25 - 0.30 & $(1.09\,^{+0.07}_{-0.10}) \times 10^{5}$ & $(1.89\,^{+0.12}_{-0.17}) \times 10^{5}$ & $(4.83\,^{+0.30}_{-0.35}) \times 10^{5}$  \\
0.30 - 0.35 & $(7.52\,^{+0.47}_{-0.85}) \times 10^{4}$ & $(1.28\,^{+0.08}_{-0.12}) \times 10^{5}$ & $(3.51\,^{+0.22}_{-0.27}) \times 10^{5}$  \\
0.35 - 0.40 & $(5.35\,^{+0.34}_{-0.59}) \times 10^{4}$ & $(8.97\,^{+0.58}_{-0.74}) \times 10^{4}$ & $(2.38\,^{+0.16}_{-0.17}) \times 10^{5}$  \\
0.40 - 0.45 & $(3.76\,^{+0.25}_{-0.39}) \times 10^{4}$ & $(6.31\,^{+0.42}_{-0.50}) \times 10^{4}$ & $(1.70\,^{+0.12}_{-0.13}) \times 10^{5}$  \\
0.45 - 0.50 & $(2.78\,^{+0.19}_{-0.24}) \times 10^{4}$ & $(4.61\,^{+0.31}_{-0.34}) \times 10^{4}$ & $(1.17\,^{+0.09}_{-0.09}) \times 10^{5}$  \\
0.50 - 0.55 & $(1.78\,^{+0.14}_{-0.18}) \times 10^{4}$ & $(3.02\,^{+0.23}_{-0.24}) \times 10^{4}$ & $(7.97\,^{+0.87}_{-0.91}) \times 10^{4}$  \\
0.55 - 0.60 & $(1.08\,^{+0.08}_{-0.10}) \times 10^{4}$ & $(1.83\,^{+0.15}_{-0.17}) \times 10^{4}$ & $(4.85\,^{+0.66}_{-0.66}) \times 10^{4}$  \\
0.60 - 0.65 & $(6.02\,^{+0.64}_{-0.65}) \times 10^{3}$ & $(1.04\,^{+0.10}_{-0.12}) \times 10^{4}$ & $(2.65\,^{+0.36}_{-0.39}) \times 10^{4}$  \\
0.65 - 0.70 & $(3.07\,^{+0.44}_{-0.44}) \times 10^{3}$ & $(5.46\,^{+0.72}_{-0.92}) \times 10^{3}$ & $(1.41\,^{+0.23}_{-0.27}) \times 10^{4}$  \\
0.70 - 0.75 & $(1.74\,^{+0.30}_{-0.34}) \times 10^{3}$ & $(2.85\,^{+0.70}_{-0.68}) \times 10^{3}$ & $(4.51\,^{+1.01}_{-1.15}) \times 10^{3}$  \\
0.75 - 0.80 & $(8.07\,^{+2.35}_{-2.22}) \times 10^{2}$ & $(1.17\,^{+0.44}_{-0.49}) \times 10^{3}$ & $(2.10\,^{+0.73}_{-0.76}) \times 10^{3}$  \\
0.80 - 0.85 & $(2.32\,^{+1.19}_{-0.88}) \times 10^{2}$ & $(2.79\,^{+1.45}_{-2.01}) \times 10^{2}$ & $(5.58\,^{+3.00}_{-2.91}) \times 10^{2}$  \\
0.85 - 0.90 & $(8.05\,^{+5.24}_{-5.01}) \times 10^{1}$ & $(8.62\,^{+5.80}_{-8.74}) \times 10^{1}$ & $(2.03\,^{+1.51}_{-1.62}) \times 10^{2}$  \\
0.90 - 0.95 & $(1.01\,^{+0.92}_{-1.08}) \times 10^{1}$ & $(1.76\,^{+1.70}_{-2.43}) \times 10^{1}$ & $(2.91\,^{+2.94}_{-3.50}) \times 10^{1}$  \\
    \hline
  \end{tabular}
\end{table}

\begin{table}
  \caption{ Differential photon production cross-section $d\,\sigma_{\gamma}/d\,x_{\mathrm{F}}$ [nb] for each \xF~bin in the pseudorapidity ranges: \etaregion{7.0}{7.5}, \etaregion{6.5}{7.0}  and \etaregion{6.1}{6.5}. }
  \label{tab:result2}
  \centering
  \begin{tabular}{cccc}
    \hline
     & \multicolumn{3}{c}{ $d\,\sigma_{\gamma}/d\,x_{\mathrm{F}}$ [nb]} \\ \cline{2-4}
    \xF  & \etaregion{7.0}{7.5} &  \etaregion{6.5}{7.0} &  \etaregion{6.1}{6.5} \\
        \hline
    0.10 - 0.15 & $(3.32\,^{+0.22}_{-0.21}) \times 10^{6}$ & $(8.24\,^{+0.51}_{-0.51}) \times 10^{6}$ & $(1.34\,^{+0.08}_{-0.08}) \times 10^{7}$  \\
0.15 - 0.20 & $(2.46\,^{+0.15}_{-0.14}) \times 10^{6}$ & $(5.91\,^{+0.35}_{-0.33}) \times 10^{6}$ & $(8.81\,^{+0.50}_{-0.49}) \times 10^{6}$  \\
0.20 - 0.25 & $(1.82\,^{+0.11}_{-0.11}) \times 10^{6}$ & $(4.20\,^{+0.26}_{-0.23}) \times 10^{6}$ & $(5.69\,^{+0.35}_{-0.31}) \times 10^{6}$  \\
0.25 - 0.30 & $(1.31\,^{+0.08}_{-0.08}) \times 10^{6}$ & $(2.79\,^{+0.18}_{-0.16}) \times 10^{6}$ & $(3.54\,^{+0.23}_{-0.20}) \times 10^{6}$  \\
0.30 - 0.35 & $(9.05\,^{+0.62}_{-0.57}) \times 10^{5}$ & $(1.85\,^{+0.13}_{-0.11}) \times 10^{6}$ & $(2.21\,^{+0.15}_{-0.13}) \times 10^{6}$  \\
0.35 - 0.40 & $(6.17\,^{+0.45}_{-0.41}) \times 10^{5}$ & $(1.22\,^{+0.09}_{-0.08}) \times 10^{6}$ & $(1.34\,^{+0.09}_{-0.08}) \times 10^{6}$  \\
0.40 - 0.45 & $(4.07\,^{+0.32}_{-0.29}) \times 10^{5}$ & $(8.02\,^{+0.61}_{-0.55}) \times 10^{5}$ & $(8.16\,^{+0.60}_{-0.54}) \times 10^{5}$  \\
0.45 - 0.50 & $(2.75\,^{+0.22}_{-0.19}) \times 10^{5}$ & $(4.91\,^{+0.39}_{-0.37}) \times 10^{5}$ & $(4.66\,^{+0.35}_{-0.33}) \times 10^{5}$  \\
0.50 - 0.55 & $(1.78\,^{+0.18}_{-0.14}) \times 10^{5}$ & $(3.02\,^{+0.25}_{-0.25}) \times 10^{5}$ & $(2.58\,^{+0.21}_{-0.20}) \times 10^{5}$  \\
0.55 - 0.60 & $(1.06\,^{+0.12}_{-0.10}) \times 10^{5}$ & $(1.68\,^{+0.16}_{-0.17}) \times 10^{5}$ & $(1.37\,^{+0.12}_{-0.12}) \times 10^{5}$  \\
0.60 - 0.65 & $(5.62\,^{+0.73}_{-0.66}) \times 10^{4}$ & $(8.68\,^{+1.03}_{-1.15}) \times 10^{4}$ & $(6.71\,^{+0.68}_{-0.78}) \times 10^{4}$  \\
0.65 - 0.70 & $(2.87\,^{+0.50}_{-0.47}) \times 10^{4}$ & $(4.28\,^{+0.62}_{-0.75}) \times 10^{4}$ & $(3.18\,^{+0.38}_{-0.49}) \times 10^{4}$  \\
0.70 - 0.75 & $(1.25\,^{+0.32}_{-0.31}) \times 10^{4}$ & $(1.99\,^{+0.34}_{-0.45}) \times 10^{4}$ & $(1.43\,^{+0.22}_{-0.30}) \times 10^{4}$  \\
0.75 - 0.80 & $(5.87\,^{+2.05}_{-2.12}) \times 10^{3}$ & $(8.47\,^{+1.74}_{-2.58}) \times 10^{3}$ & $(5.83\,^{+1.13}_{-1.66}) \times 10^{3}$  \\
0.80 - 0.85 & $(2.98\,^{+1.40}_{-1.57}) \times 10^{3}$ & $(2.68\,^{+0.67}_{-1.10}) \times 10^{3}$ & $(1.65\,^{+0.40}_{-0.65}) \times 10^{3}$  \\
0.85 - 0.90 & $(1.05\,^{+0.69}_{-0.77}) \times 10^{3}$ & $(7.23\,^{+3.32}_{-4.89}) \times 10^{2}$ & $(4.85\,^{+2.08}_{-2.83}) \times 10^{2}$  \\
0.90 - 0.95 & $(4.50\,^{+4.24}_{-4.69}) \times 10^{1}$ & $(8.99\,^{+7.97}_{-10.41}) \times 10^{1}$ & $(8.58\,^{+7.69}_{-8.62}) \times 10^{1}$  \\
    \hline
  \end{tabular}
\end{table}

\begin{table}
    \caption{ 
    Production cross section ($\Delta \sigma_{\gamma,x_{\mathrm{F}}>0.1}$) and energy flow ($\Delta E_{\gamma,x_{\mathrm{F}}>0.1}$) of forward photons with \xF~$>$ 0.1 in each pseudorapidity ranges. Due to avoiding the artificial normalization in the largest rapidity bin, these values were computed without the normalization by the width of each pseudorapidity bin $\Delta \eta$ in Eq.~\ref{eq-integralsigma} and \ref{eq-integralflux}. The values in the right column show the fraction of energy flow contributed from photons with \xF~$>$ 0.1 over that from all photons $\frac{\Delta E_{\gamma,x_{\mathrm{F}}>0.1}}{\Delta E_{\gamma,x_{\mathrm{F}}>0}}$ estimated by \DPMJET. }
    \label{tab:result3}
    \centering
    \begin{tabular}{cccc}
      \hline
      $\eta$ & $\Delta\sigma_{\gamma,x_{\mathrm{F}}>0.1}$ [mb] &
        $\Delta E_{\gamma,x_{\mathrm{F}}>0.1}$ [GeV] & 
        $\frac{\Delta E_{\gamma,x_{\mathrm{F}}>0.1}}{\Delta E_{\gamma,x_{\mathrm{F}}>0}}$ (\DPMJET)\\
          \hline
6.1 - 6.5 & $1.84\,^{+0.10}_{-0.11}$ & $1.97\,^{+0.13}_{-0.13}$ & 79.6\% \\
6.5 - 7.0 & $1.30\,^{+0.08}_{-0.08}$ & $1.51\,^{+0.10}_{-0.10}$ & 83.7\% \\
7.0 - 7.5 & $0.58\,^{+0.04}_{-0.04}$ & $0.71\,^{+0.05}_{-0.05}$ & 86.0\% \\
7.5 - 8.0 & $0.214\,^{+0.014}_{-0.015}$ & $0.266\,^{+0.019}_{-0.021}$ & 86.9\% \\
8.0 - 8.5 & $0.080\,^{+0.005}_{-0.006}$ & $0.100\,^{+0.007}_{-0.008}$ & 87.3\% \\
$>$ 8.5 & $0.0467\,^{+0.0030}_{-0.0038}$ & $0.0590\,^{+0.0041}_{-0.0053}$ & 87.4\% \\
      \hline
    \end{tabular}
  \end{table}

\afterpage{\clearpage}
\newpage

 \bibliographystyle{elsarticle-num} 
 \bibliography{rhicfphoton}





\end{document}